\documentstyle[multicol,aps,prl,psfig]{revtex} 

\renewcommand{\narrowtext}{\begin{multicols}{2} \global\columnwidth20.5pc}
\renewcommand{\widetext}{\end{multicols} \global\columnwidth42.5pc}
\def\al{\alpha}
\def\be{\beta}
\def\ga{\gamma}
\def\de{\delta}

\def\et{\eta}
\def\th{\theta}

\def\la{\lambda}

\def\ch{\chi}
\def\ps{\psi}

\def\Om{\Omega}

\def\cL{{\cal L}}

\def\fr#1#2{{{#1} \over {#2}}}
\def\frac#1#2{\textstyle{{{#1} \over {#2}}}}
\def\pt#1{\phantom{#1}}
\def\prt{\partial}
\def\vev#1{\langle {#1}\rangle}
\def\ket#1{|{#1}\rangle}

\def\half{{\textstyle{1\over 2}}}
\def\lsim{\mathrel{\rlap{\lower4pt\hbox{\hskip1pt$\sim$}}
    \raise1pt\hbox{$<$}}}
\def\gsim{\mathrel{\rlap{\lower4pt\hbox{\hskip1pt$\sim$}}
    \raise1pt\hbox{$>$}}}

\def\lrDmu{\stackrel{\leftrightarrow}{D_\mu}}
\def\lrDnu{\stackrel{\leftrightarrow}{D^\nu}}
\def\lrDbe{\stackrel{\leftrightarrow}{D_\be}}
\def\lrhatDmu{\stackrel{\leftrightarrow}{\widehat D_\mu}}

\newcommand{\beq}{\begin{equation}}
\newcommand{\eeq}{\end{equation}}
\newcommand{\bea}{\begin{eqnarray}}
\newcommand{\eea}{\end{eqnarray}}
\newcommand{\rf}[1]{(\ref{#1})}

\begin{document}

\title{Noncommutative Field Theory and Lorentz Violation} 
\author{
Sean M.\ Carroll,$^a$
Jeffrey A.\ Harvey,$^a$
V.\ Alan Kosteleck\'y,$^b$
Charles D.\ Lane,$^c$
and Takemi Okamoto$^a$}

\address{
$^a$Enrico Fermi Institute,
University of Chicago,
Chicago, IL 60637, U.S.A.}
\address{
$^b$Physics Department, Indiana University, 
Bloomington, IN 47405, U.S.A.}
\address{
$^c$Physics Department, Colby College, 
Waterville, ME 04901, U.S.A.}

\date{EFI 01-12, IUHET 433, May 2001} 

\maketitle

\begin{abstract}
The role of Lorentz symmetry in noncommutative field theory
is considered.
Any realistic noncommutative theory is found to be physically equivalent
to a subset of a general Lorentz-violating standard-model extension
involving ordinary fields.
Some theoretical consequences are discussed.
Existing experiments bound the scale of the noncommutativity parameter 
to (10 TeV)$^{-2}$.
\end{abstract}


\narrowtext

The idea that spacetime may intrinsically involve
noncommutative coordinates has undergone a recent revival 
following the realization that this occurs 
naturally in string theory
\cite{cds}.
In this framework,
the commutator of the coordinates $x^\mu$ 
in the spacetime manifold is:
\beq
[ x^\mu , x^\nu ] = i\th^{\mu\nu},
\label{comm}
\eeq
where $\th^{\mu\nu}$ is real and antisymmetric.
It is of interest to speculate that the physical world 
might involve noncommutative coordinates
and to ask about current experimental 
sensitivity to putative realistic noncommutative 
quantum field theories.

The primary goal of this work is to study a physical issue
that is central to any realistic noncommutative theory: 
the role of Lorentz symmetry.
Violations of Lorentz symmetry are intrinsic to
noncommutative theories by virtue of nonzero $\th^{\mu\nu}$
in Eq.\ \rf{comm}.
Our study of these violations is motivated partly 
by theoretical progress in understanding the physics
associated with Lorentz violation in ordinary quantum field theory
and partly by recent experimental advances
that make Lorentz tests among the most sensitive 
null experiments in existence
\cite{cpt98}.

One approach to constructing a 
noncommutative quantum field theory
is to promote an established ordinary theory 
to a noncommutative one  
by replacing ordinary fields with noncommutative fields
and ordinary products with Moyal $\star$ products,
defined by
\beq
f\star g(x) \equiv
\exp(\half i\th^{\mu\nu}\prt_{x^\mu}\prt_{y^\nu}) f(x) g(y)\big|_{x=y}.
\label{moyal}
\eeq
For gauge theories
such as quantum electrodynamics (QED),
ordinary gauge transformations must be modified  
to noncommutative generalizations.
For noncommutative QED
\cite{mh},
the hermitian lagrangian is
\beq
\cL = 
\half i \overline{\widehat{\ps}} \star
\ga^\mu \lrhatDmu \widehat \ps 
- m \overline{\widehat{\ps}} \star \widehat \ps 
- \fr 1 {4q^2} \widehat F_{\mu\nu} \star \widehat F^{\mu\nu}.
\label{ncqed}
\eeq
Here,
carets indicate noncommutative quantities,
$\widehat F_{\mu\nu} 
= \prt_\mu\widehat A_\nu
-\prt_\nu\widehat A_\mu
- i [\widehat A_\mu, \widehat A_\nu]_\star$,
and $\widehat D_\mu \widehat\ps 
= \prt_\mu\widehat\ps - i \widehat A_\mu \star \widehat\ps$,
with 
$\widehat f \star \lrhatDmu \widehat g \equiv
\widehat f \star \widehat D_\mu \widehat g 
- \widehat D_\mu \widehat f \star \widehat g$.
Note that the inclusion of particles of charge other than 0 or $\pm 1$
is problematic
\cite{mh}.
This poses difficulties for
a noncommutative generalization of the standard model,
which would require other values for hypercharge assignments.
In fact,
noncommutative QED is similar 
to U($N$) gauge theory as $N \rightarrow \infty$,
and the allowed representations are the adjoint,
fundamental, and antifundamental. 
In D-brane physics,
adding two D-branes of charge 1 under a noncommutative U(1)
leads to noncommutative U(2) gauge theory,
which has nonabelian U(2) gauge theory as its commutative limit 
instead of U(1) with charge 2. 

The implementation of Lorentz transformations in 
a noncommutative theory 
is more involved than usual 
because the parameter $\th^{\mu\nu}$ carries Lorentz indices.
Two distinct types of Lorentz transformation exist
\cite{ck}.
For example,
Eq.\ \rf{ncqed} is fully covariant under 
observer Lorentz transformations:
rotations or boosts of the observer inertial frame
leave the physics unchanged 
because both the field operators and $\th^{\mu\nu}$
transform covariantly.
However,
these coordinate changes differ profoundly
from rotations or boosts of a particle or localized field configuration
within a fixed observer frame.
The latter, 
called particle Lorentz transformations,
leave $\th^{\mu\nu}$ unaffected
and hence modify the physics.
This situation is closely analogous to the result of
spontaneous Lorentz violation
\cite{kps},
with $\th^{\mu\nu}$ playing the role of
a tensor expectation value.
In effect,
$\th^{\mu\nu}$ provides a 4-dimensional directionality to spacetime
in any fixed inertial frame.
Any noncommutative theory
therefore violates particle Lorentz symmetry.

The procedure leading to Eq.\ \rf{ncqed} 
lacks direct information about
the identification of realistic physical variables
with specific operators.
For instance,
the electron field $\widehat\ps$ in the noncommutative QED \rf{ncqed}
is itself noncommutative
and obeys an unconventional gauge transformation law,
so the identification of its quantum with the physical electron 
is nontrivial.
Although it is presumably feasible in principle to 
calculate physical observables via noncommutative fields,
we use here instead a correspondence between
a noncommutative gauge theory and a conventional gauge theory,
called the Seiberg-Witten map 
\cite{sw}.
This permits the construction of an ordinary theory
with ordinary gauge transformations
having physical content guaranteed equivalent 
to the noncommutative theory.

The existence of an equivalent ordinary gauge theory
for any realistic noncommutative theory
involving noncommutative standard-model fields
is of interest because there already exists
a general extension of the ordinary standard model 
allowing for Lorentz violation
\cite{kp,ck}.
This theory can be defined as the standard model 
lagrangian plus all possible gauge-invariant terms 
involving standard-model fields
that preserve observer Lorentz invariance
while breaking particle Lorentz symmetry.
It therefore follows that
\it 
any realistic noncommutative theory 
must be physically equivalent to
a subset of the standard-model extension.
\rm

A variety of theoretical and experimental implications
of the standard-model extension are known,
and the existence of the equivalence ensures 
some of these also hold for any realistic noncommutative theory.
The Lorentz-violating terms in the standard-model extension
are contractions of field operators that transform
as Lorentz tensors with coefficients that carry observer Lorentz indices.
In any subset of this theory 
equivalent to a noncommutative theory,
the coefficients for Lorentz violation
must be expressed solely in terms of $\th^{\mu\nu}$.
This has several immediate consequences for
any realistic noncommutative theory.
As a simple example,
energy and momentum are conserved in the full standard-model extension
provided the coefficients for Lorentz violation are constant.
Since $\th^{\mu\nu}$ is independent of spacetime position,
this condition is satisfied and so 
\it energy and momentum are conserved 
in any realistic noncommutative theory.
\rm 

As another example,
terms in the standard-model extension violate CPT 
if and only if the coefficients for Lorentz violation  
carry an odd number of indices.
Since it is impossible to construct such a coefficient
from combinations of $\th^{\mu\nu}$,
it immediately follows that 
\it
any realistic noncommutative theory necessarily preserves CPT.
\rm
This generalizes a result obtained for the case of noncommutative QED
\cite{ms}.
In contrast,
all other combinations of the discrete symmetries C, P, T 
can be broken in a general noncommutative theory.

Further insight is provided by the observation 
that in a noncommutative field theory 
each factor of $\th^{\mu\nu}$ is accompanied by two derivatives.
Since bilinear fermion operators in a noncommutative theory
have mass dimension 3 or 4,
the minimal dimension 
of the corresponding Lorentz-violating bilinear operators 
in the equivalent lagrangian is 5 or 6.
In fact,
higher-dimensional terms and nonlocal interactions
are required for consistency at high scales 
in the full standard-model extension 
\cite{kle}.
However,
the absence here of Lorentz-violating operators 
of dimension 3 or 4 implies 
\it
the fermionic sector of any realistic noncommutative theory
is free of perturbative difficulties with stability and causality.
\rm
This implies,
for example,
the absence of superluminal information transfer.

Some noncommutative theories with $\th^{0j}\neq 0$ 
exhibit difficulties with perturbative unitarity
\cite{gm},
but ones with only $\th^{jk}$ nonzero are acceptable.
Since a theory with $\th^{0j} \neq 0$ and $\th_{\mu\nu}\th^{\mu\nu}>0$
can be converted into one with only $\th^{jk}$ nonzero
by a suitable observer Lorentz transformation,
the presence of observer Lorentz invariance implies that
\it 
there are no difficulties with perturbative unitarity 
provided $\th_{\mu\nu}\th^{\mu\nu}>0$,
\rm
which allows certain cases with $\th^{0j} \neq 0$.
A similar condition presumably applies for open bosonic strings,
where the presence of a nonzero $B^{jk}$ field
is known to be equivalent to a constant magnetic field on a D$p$-brane
\cite{sw}.
In the standard-model extension,
Lorentz-violating operators with extra time-derivative couplings
do cause some interpretational difficulties,
but these can be handled by redefining the fields
to evolve canonically
\cite{bkr,kle}.
We expect analogous methods 
to apply for noncommutative theories
with $\th^{0j} \neq 0$.

For definiteness,
we focus primarily on the noncommutative QED \rf{ncqed}
with $\th_{\mu\nu}\th^{\mu\nu}>0$
in the remainder of this work.
In this case,
the explicit form of the Seiberg-Witten map is known
to lowest order in $\th^{\mu\nu}$
\cite{sw,bgpsw}:
\bea
\widehat A_\mu &=& 
A_\mu - \half \th^{\al\be} A_\al
(\prt_\be A_\mu + F_{\be\mu}),
\nonumber\\
\widehat\ps &=& \ps - \half \th^{\al\be}
A_\al \prt_\be \ps.
\label{swmap}
\eea
This leading-order form suffices for many purposes,
since any physical noncommutativity in nature must be small.

Substitution of the solution \rf{swmap}
into Eq.\ \rf{ncqed} 
and applying the definition \rf{moyal}
yields the ordinary quantum field theory 
that is physically equivalent to noncommutative QED
to leading order in $\th^{\mu\nu}$:
\bea
\cL &=& 
\half i \overline{\ps} \ga^\mu \lrDmu \ps
- m \overline{\ps} \ps
- \frac 1 4 F_{\mu\nu} F^{\mu\nu}
\nonumber\\
&&
- \frac 1 8 i q\th^{\al\be} F_{\al\be}
\overline{\ps} \ga^\mu \lrDmu \ps
+ \frac 1 4 i q\th^{\al\be} F_{\al\mu}
\overline{\ps} \ga^\mu \lrDbe \ps
\nonumber\\
&&
+ \frac 1 4 m q \th^{\al\be}F_{\al\be} \overline{\ps} \ps
\nonumber\\
&&
- \frac 1 2 q \th^{\al\be} F_{\al\mu} F_{\be\nu} F^{\mu\nu}
+ \frac 1 8 q \th^{\al\be} F_{\al\be} F_{\mu\nu} F^{\mu\nu}.
\label{oqed}
\eea
In this equation,
we have redefined the gauge field $A_\mu \to q A_\mu$
to display the charge coupling of the physical fermion,
and $D_\mu \ps = \prt_\mu\ps - i q A_\mu\ps$ as usual.

The expression \rf{oqed}
is manifestly gauge invariant.
It consists of ordinary QED plus 
nonrenormalizable Lorentz-violating corrections
and is therefore a subset of the QED limit of
the standard-model extension,
as expected.
However,
many terms allowed
in the latter theory are absent,
including all those that violate CPT.
Note that the $\ga$-matrix structure in Eq.\ \rf{oqed}
is inherited from the usual one in Eq.\ \rf{ncqed}, 
so no couplings to axial-vector or tensor bilinears appear.
Note also that all noncommutative effects vanish 
for neutral fermions.

With this explicit theory in hand,
we can consider some possible experimental implications
of noncommutativity.
Here,
we focus attention on the case of experiments
involving constant electromagnetic fields.
For this purpose,
it is useful to extract from the theory \rf{oqed} 
an effective lagrangian 
describing the leading-order effects of noncommutativity
in constant electromagnetic fields.
We therefore make the replacement  
$F_{\mu\nu} \to  f_{\mu\nu} + F_{\mu\nu}$,
where $f_{\mu\nu}$ is understood to be a constant background field
and $F_{\mu\nu}$ now denotes a small dynamical fluctuation.

Keeping only terms up to quadratic order in the fluctuations
and performing a physically irrelevant rescaling
of the fields $\ps$ and $A_\mu$
to maintain conventionally normalized kinetic terms
yields the hermitian lagrangian
\bea
\cL &=& 
\half i \overline{\ps} \ga^\mu \lrDmu \ps
- m\overline{\ps} \ps
- \frac 1 4 F_{\mu\nu} F^{\mu\nu}
\nonumber\\
&&
+ \half i c_{\mu\nu} \overline{\ps} \ga^\mu \lrDnu \ps
- \frac 1 4 {k_F}_{\al\be\ga\de} F^{\al\be} F^{\ga\de}.
\label{fqed}
\eea
In this equation,
the charge $q$ in the covariant derivative is replaced
with a scaled effective value
\cite{fn1}
\beq
q_{\rm eff} = (1 + \frac 1 4 q f^{\mu\nu}\th_{\mu\nu}) q.
\label{charge}
\eeq
The coefficients $c_{\mu\nu}$ and ${k_F}_{\al\be\ga\de}$ are
\bea
c_{\mu\nu} &=& - \half q f_\mu^{\pt{\mu}\la} \th_{\la\nu},
\nonumber\\
{k_F}_{\al\be\ga\de} &=& 
- q f_\al^{\pt{\al}\la} \th_{\la\ga}\et_{\be\de}
+ \half q f_{\al\ga} \th_{\be\de} 
- \frac 1 4 q f_{\al\be} \th_{\ga\de}
\nonumber\\
&&
- (\al \leftrightarrow \be)
- (\ga \leftrightarrow \de)
+ (\al\be \leftrightarrow \ga\de).
\label{coeffs}
\eea
The notation here
is that of the standard-model extension in its QED limit
\cite{ck}.
Of the 10 types of term allowed in the latter theory,
six are excluded here by CPT symmetry
and two by the requirement of no couplings to axial or tensor
fermion bilinears.
However,
some caution is required in applications 
because the coefficients
$c_{\mu\nu}$ and 
${k_F}_{\al\be\ga\de}$
now depend on the background field strength.

In the photon sector,
there are presently no published bounds on the coefficients
${k_F}_{\al\be\ga\de}$.
The modified Maxwell equations \it in vacuo \rm
have been studied,
and it appears feasible to place bounds at the scale
of about $10^{-28}$ on certain components of 
${k_F}_{\al\be\ga\de}$,
using measurements of the birefringence of radiation
from cosmological sources
\cite{ck,cfj}.
However,
the dependence of ${k_F}_{\al\be\ga\de}$
on the minuscule intergalactic magnetic field
and the likely dilution of any effect
due to random field orientations
implies only weak bounds on $\th^{\mu\nu}$ are likely.

Instead,
we turn to the fermion sector.
Numerous tests of Lorentz violation have been performed 
in the context of the standard-model extension,
but many of them can detect only CPT violation
or anomalous spin couplings
and so place no bounds on $c_{\mu\nu}$. 
One class of tests with sensitivity to $c_{\mu\nu}$
involves the recent clock-comparison experiments
\cite{ccexpt}.
These monitor the difference 
between two atomic hyperfine or Zeeman transition frequencies,
searching for variations as the Earth rotates.
The existing analysis \cite{kla}
of the implications of these experiments
can be adapted to the present situation. 

The energy shift $\de$ in an atomic state 
labeled $\ket{F, m_F}$
can be calculated as the expectation value 
of the hermitian perturbation hamiltonian 
obtained from Eq.\ \rf{fqed}.
It has the form
$\de \sim 
\tilde m_{F} \sum_w \ga_w \tilde c_q^w$,
where $\tilde m_F$ is a ratio of Clebsch-Gordan coefficients,
$w$ labels the particle species (electron, proton, neutron) 
of mass $m_w$ and charge $q_w$,
$\ga_w$ is an expectation value of momentum operators 
in an extremal state of the atomic or nuclear submanifold of levels,
and $\tilde c_q^w \equiv m_w(c_{11} + c_{22} - 2 c_{33})$ 
is a quadrupole combination of coefficients for Lorentz violation.
Expressions for $\tilde m_F$ and $\ga_w$
are provided in Eqs.\ (7) and (10) of Ref.\ \cite{kla}.
With the magnetic field $F^{12}\equiv - B$
along the $3$ axis in the laboratory frame,
we find $c_q^w = m_w q_w B \th^{12}$.

In the laboratory frame rotating with the Earth,
the parameters $m_w$, $q_w$, and $B$ are fixed 
but $\th^{12}$ varies with time $t$. 
To display the corresponding $t$ dependence of the energy shift $\de$,
we use a nonrelativistic transformation 
to convert to nonrotating coordinates $(X,Y,Z)$
compatible with celestial equatorial coordinates
\cite{kla}. This gives
\beq
\de = E_0 + E_{1X}\cos\Om t + E_{1Y}\sin\Om t,
\eeq
where $\Om$ is the Earth's sidereral rotation frequency,
$E_0$ is an irrelevant constant,
and
\beq
(E_{1X}, E_{1Y})
= \tilde m_FB\sin{\ch}
\sum_{w} q_w m_w \ga_w (\th_{YZ}, \th_{ZX}),
\label{ens}
\eeq
with $\ch$ the angle between the $3$ and the $Z$ axes.
Note that,
despite its quadrupole nature,
the energy shift $\de$ is periodic in $\Om t$.
This contrasts with the $2 \Om t$ dependence arising
when $c_{\mu\nu}$ is independent 
of $B$.

We can apply these results to recent clock-comparison tests
\cite{ccexpt}.
Most place bounds on variations with $\Om t$
but in the context of the Schmidt nuclear model
\cite{ts}
are sensitive only to effects from the neutron,
which vanish here 
because the neutron is uncharged.
The exception is the experiment
of Berglund
{\it et al.},
which compares transitions in $^{199}$Hg and $^{133}$Cs.
In this experiment,
as in the others,
the electronic angular momentum is $J\leq 1/2$,
so $\ga_e$ vanishes and there is no signal
associated with the electron.
However,
the nuclear spin of the $^{133}$Cs atom is $I=7/2$,
and the Schmidt model predicts the valence nucleon to be a proton,
so the experimental limit of about 100 nHz 
on possible sidereal variations 
yields a bound
$|\th^{YZ}|$, $|\th^{ZX}| \lsim$ (10 GeV)$^{-2}$.

The above bound is suppressed due to the weak magnetic field 
($B\sim 5$ mG) used in the experiment.
In contrast, 
the experiment of Prestage {\it et al.}
involves an applied field of about 1 T.
It is therefore worth considering 
possible effects outside the Schmidt model.
The explicit value of $\ga_p$ in Eq.\ \rf{ens}
is an expectation value of momentum operators
in the multiparticle wave function $\ket{\ps}$
for the $^9$Be nucleus used in the experiment
\cite{rb}:
\beq
\ket{\ps} = 
C_1 ({^1S}, {^2P}) + C_2 ({^1D}, {^2P}) + C_3 ({^1D}, {^2D}).
\label{wvfn}
\eeq
Each term in parentheses refers to the proton and neutron
spin and orbital angular momentum 
according to $({^{2S_p+1}L_p}, {^{2S_n+1}L_n})$,
and the coefficients are 
$C_1 \simeq 0.731$,
$C_2 \simeq -0.344$,
$C_3 \simeq -0.589$.
A calculation gives
\beq
\ga_p = - [ 7(C_2^2 - C_3^2) + 8 \sqrt 5 C_1 C_2] K_p/150.
\label{est}
\eeq
Here,
$K_p = \vev{p^2}_p/m_p^2 \simeq 10^{-2}$,
yielding $\ga_p\simeq 5 \times 10^{-4}$.
Using the results of Prestage {\it et al.},
we find
\beq
|\th^{YZ}|$, $|\th^{ZX}| \lsim (10 {\rm ~TeV})^{-2}
\label{bound}
\eeq
as a conservative limit
\cite{fn2}.

Other low-energy bounds on $\th^{\mu\nu}$ exist. 
Measurements of the Lamb shift give
\cite{cst}
a bound several orders of magnitude weaker than \rf{bound}.
A speculative bound some 20 orders of magnitude 
stronger than \rf{bound} has been claimed 
\cite{mpr}
from an analysis of clock-comparison experiments.
This analysis finds terms with anomalous spin couplings
and obtains a bound by supposing that,
in an eventual formulation of noncommutative quantum chromodynamics,
such couplings would produce a coherent effect involving the nuclear force.

For bounds on $c_{\mu\nu}$,
high-energy experiments appear to provide no particular advantage
over low-energy ones,
basically because the effects scale 
with momentum like those from the usual fermion kinetic term
\cite{ck2}.
Assuming the interactions in Eq.\ \rf{oqed}
affect at least some high-energy cross sections, 
the attainable high-energy bound
can be crudely estimated as about (1 TeV)$^{-2}$
by noting that leading-order couplings involving $\th^{\mu\nu}$ 
come with two powers of momentum,
while cross sections at 100 GeV are typically 
known to no better than about 1\%.
This bound is compatible with existing analyses 
\cite{hpr}.

Further theoretical analysis
might improve the bound \rf{bound}.
For example,
it may be worth studying 
the effect of the magnetic field at the nucleus
caused by atomic electrons,
since under suitable circumstances 
the effect of this field in $c_{\mu\nu}$
might dominate the applied one. 
Also,
additional experimental sensitivity might arise if
the neutron is coupled in the adjoint representation of 
noncommutative QED.
The range of relevant tests might be further broadened
if additional $\ga$-matrix structures arise
in radiative corrections in the theory \rf{oqed} 
or in more complicated versions of noncommutative QED
obtained from 
the radiative effective action in ordinary QED.

Several experimental options exist 
for improving the bound \rf{bound}.
One would be to perform 
a clock-comparison test in a large field 
using substances that are particularly favorable
for theoretical calculations.
These include the subset of species listed in
Table III of Ref.\ \cite{kla}
that have quadrupole sensitivity to proton effects.
It would be ideal to compare one such species
to a reference 
for which noncommutative effects are absent.
For example,
an experiment comparing transitions in 
$^{209}$Bi with $^3$He
or,
perhaps more feasibly,
$^{87}$Rb with $^3$He
has the potential to provide an improved reliable bound
on $\th^{\mu\nu}$.

\medskip

This work was supported in part 
by the Department of Energy,
the National Science Foundation,
the Packard Foundation,
and the Sloan Foundation.

\end{multicols}

\begin{references}
\bibitem{fn1}
The effective charge varies with the applied field.
In a 1 T magnetic field with the maximum possible $\th^{jk}$,
$q_{\rm eff}$ differs by only about a part in $10^{24}$
from its zero-field value,
which is presently unobservable.
\bibitem{fn2}
Using an exact relativistic transformation
between the laboratory and nonrotating frames
in deriving Eq.\ \rf{ens}
reveals contributions from components $\th^{0J}$
that are suppressed by factors of about $10^{-6}$.
We therefore also obtain 
bounds of roughly (10 GeV)$^{-2}$ on these components,
although an accurate treatment requires incorporation
of subleading effects disregarded here.
\end{references}
\end{document}